\documentclass[12pt,superscriptaddress]{revtex4}
\usepackage{graphics,graphicx}
\input epsf

\begin{document}
\baselineskip 14pt
\title{Another Look at Quantum Teleportation}

\author{L. Vaidman}
\affiliation{ \baselineskip 13pt School of Physics and Astronomy\\
Raymond and Beverly Sackler Faculty of Exact Sciences \\
Tel-Aviv University, Tel-Aviv 69978, Israel}

\author{N. Erez}
\affiliation{ \baselineskip 13pt School of Physics and Astronomy\\
Raymond and Beverly Sackler Faculty of Exact Sciences \\
Tel-Aviv University, Tel-Aviv 69978, Israel}

\affiliation{ \baselineskip 13pt Institute for Quantum Studies and
Physics Dept., Texas A\&M University, College Station, TX
77843-4242}

\author{A. Retzker}
\affiliation{ \baselineskip 13pt School of Physics and Astronomy\\
Raymond and Beverly Sackler Faculty of Exact Sciences \\
Tel-Aviv University, Tel-Aviv 69978, Israel}

\vspace{.4cm}

\begin{abstract}\baselineskip 14pt
 A dialog with Asher Peres regarding the meaning of quantum
 teleportation is briefly reviewed. The Braunstein-Kimble method for
 teleportation of light is analyzed in the language of quantum wave
 functions. A pictorial example  of continuous variable teleportation is
presented using computer simulation.
 \end{abstract}


\maketitle

\section{Introduction}

 I, Lev Vaidman, knew Asher Peres since the beginning of
my interest in the Foundations of Quantum Mechanics. Mostly, we were
fighting to prove that the interpretations of quantum effects
adopted by each of us were better. Asher was against the usage by
Aharonov and myself of a quantum state evolving backwards in time
\cite{AV91,Pretpar,AVcom,Prep} and contesting our interpretation of
the physical meaning of ``weak measurements"
\cite{AAV100,Pcom,AVrep}. He objected to the names I gave to my
proposals like ``cryptography with orthogonal states"
\cite{GV,PcomGV,GVrep} or ``interaction-free
measurements"\cite{Ppriv}.
 Our disagreements did not make our
interactions less fruitful: we agreed about physical facts and
discussion of the interpretation only sharpened our (at least mine)
understanding of various aspects of these effects.

Even the most basic disagreement, where Asher says: ``Quantum
mechanics needs no interpretation"\cite{Pni}, and I write that the
many-worlds interpretation is by far the best way to view quantum
mechanics \cite{myMWI}, is also essentially a disagreement only
about names. Asher explains his view describing Kathy, an
experimental physicists who, after making a quantum experiment
becomes a superposition of a lady who ate a cake and a lady who ate
a fruit. Asher says that even at this stage, in principle, it is
still possible to reverse the evolution and come back to the state
which was before the quantum measurement. For me, this story is a
gedanken test of the many-worlds interpretation. We completely agree
on the facts: There is no such thing as the collapse of the quantum
wave function. I interpret this story as splitting and reunion (with
the help of super-technology) of two worlds, while Asher, to avoid
paradoxes, considers this as an argument in favor of the approach
according to which quantum mechanics should not be taken as a
description of an objective reality.

 The main part of this paper is devoted to
the renewed analysis, performed together with two members of the
quantum group of Tel-Aviv University, of the topic on which I and
Asher, in a way, collaborated: this is the issue of teleportation. I
was very pleased to hear from Asher that finally we came to an
agreement. In the last e-mail I received from him two weeks before
he left us, he recommended to the Jerusalem Report to interview me
instead of him (due to his health condition) about teleportation. In
his paper ``What is actually teleported?" \cite{Pwat}, Asher is
joking about the suggestion of Charlie Bennett to cite the ``weak
measurements" of Aharonov and myself, but he mentions that in
another work \cite{VAA} there are seeds of the teleportation paper
\cite{bennet}.

 Indeed, I had the tools to find the solution for the teleportation
problem. When I saw the abstract of this seminal paper I immediately
new how to do it (in my way). But it took the genius of Asher and
his collaborators to ask the question. Still, my way of
teleportation was useful too. I proposed a method for two-way
teleportation which is applicable also for continuous variables
\cite{mytele}. The importance of this work became clear only a few
years later, when Braunstein and Kimble found a realistic way to
implement the continuous variables teleportation experiment using
squeezed light \cite{BK}. This experiment was successfully
accomplished in 1998 \cite{Furu1} and recently improved
 \cite{Furu2}.

Braunstein and Kimble \cite{BK} described their proposal in the
language of Wigner functions, the common approach of the quantum
optics community. Numerous analysis and further experiments since
then mostly continued to use the Wigner function formalism. I
believe that the language of quantum states has advantages in
discussions of the Foundations of Quantum mechanics, so it is of
interest to present the Braunstein-Kimble experiment in the language
of quantum states. Section 3 is devoted to this purpose. In Section
4 the results of Section 4 are demonstrated on a particular example.
But before this, I cannot resist the temptation to continue the
interpretation dialogue with Asher.

\section{ What is actually teleported?}

In the framework of classical physics, teleportation, defined as
e.g. ``theoretical transportation of matter through space by
converting it into energy and then reconverting it at the terminal
point,'' the quotation from Asher's dictionary (Webster), is
obviously a science fiction concept. Massive objects, ``matter"
 cannot ``jump"
from one place to another. The definition in The Oxford Dictionary
\begin{quotation}
\noindent  {\bf teleportation}. {\it Psychics} and {\it Science
    Fiction}.
 The conveyance
  of persons (esp. of oneself) or things by psychic power; also in
  futuristic description, apparently instantaneous transportation of
  persons, etc., across space by advanced technological
means.
\end{quotation}
sounds equally impossible for implementation. However, quantum
theory makes it more plausible. According to quantum theory, all
elementary particles of the same kind are identical. There is no
difference between the electrons in my body and the electrons in a
rock on the moon. Thus, what defines a particular person is not a
collection of the elementary particles he is made of, but {\it the
quantum state} of these particles. If I want to move to the moon, I
need not move my electrons, protons, etc. to the moon. It is enough
to reconstruct the quantum state of the same particles there. From
my point of view, I {\it am} the quantum state, so creation of this
quantum state on the moon is  my teleportation to the moon. Compare
this view with Asher's reply when he was asked by a newsman, whether
it was possible to teleport not only the body but also the soul:
``only the soul.''

In this approach, teleportation sounds as trivial as a FAX machine,
but it is not. There are two reasons why it seems impossible. First,
it is impossible to measure (to scan) the quantum state. Second, the
amount of information needed to specify a quantum state even of a
small object is so huge that it is not feasible to transmit it in a
reasonable time. The dual channel of quantum teleportation does the
trick: the quantum state is teleported without being scanned. The
quantum channel consists of entangled pairs of elementary particles,
as many as we need for the object to be teleported. Originally, at
the remote location there is a mixture of different states in which
probability for any state is the same. Then, local joint measurement
performed on the system to be teleported together with the local
part of the quantum channel specifies the particular decomposition
of the mixture in the remote location with relatively small number
of states. Finally, the only information to be transmitted is the
number of the ``actual" state in the mixture. In my view,
\cite{parad}, the local measurement creates numerous worlds with the
teleported quantum state which is deformed in various ways. The
final stage of teleportation is the correction of the deformation
such that in all worlds the final state of the remote system is the
initial state of the local system.

 Clearly, Asher would not join me
considering myself as an (unknown) quantum state. For him, a quantum
state is just the knowledge of the preparer \cite{Pajp}:
\begin{quotation}
A state vector is not a property of a physical system (nor of an
ensemble of systems). ... Rather, a state vector represents a
procedure for preparing or testing one or more physical systems.
\end{quotation}
 Then, the correction is
really not that important: the preparer knows that I am teleported
in a particular deformed way. The ``deformed me" is, probably, not a
living creature at all, so I tend not to accept Asher's approach.
But, since we are both sure that
 now and in any
foreseeable future, a realistic teleportation experiment with people
{\it is} a science fiction story, this disagreement is irrelevant.

\section{Wave Function Description of the Braunstein-Kimble Scheme}

In their seminal paper on the implementation of continuous
teleportation with squeezed light \cite{BK}, Braunstein and Kimble
had used the Wigner representation. In this section, we explain
their method using wave functions.

\begin{figure}
\begin{center}
\includegraphics[width=0.55\textwidth]{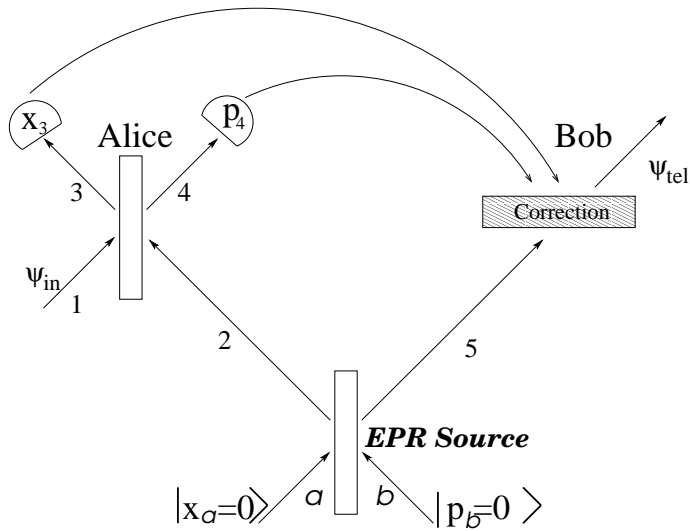}
\end{center}
\caption{The Braunstein-Kimble Teleportation scheme} \label{telepo}
\end{figure}

Figure \ref{telepo} is a schematic representation of the
experimental setup envisioned by Braunstein and Kimble. The (single
mode) state of the beam incident on the ``in port'' is to be
teleported to the ``out port''. To this end, a highly squeezed
two-mode state is used (which leaves the source marked ``EPR''). One
half of the ``EPR-pair'' is combined via a 50-50 beam splitter with
the ``in'' beam and the two resulting beams are measured using
homodyne detectors $D_x$ and $D_p$ measuring $x$ and $p$
appropriately. The results of these measurements are then used to
implement corrections on the other (remote) half of the EPR pair
which leave it in a state which closely approximates the input
state.

\begin{figure}
\begin{center}
\includegraphics[width=4.2 cm ]{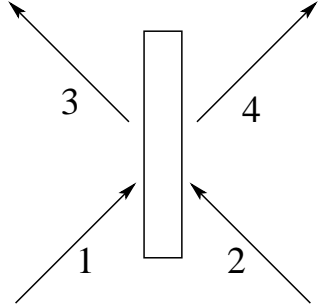}
\end{center}
\caption{Beam Splitter } \label{bs1}
\end{figure}

For simplicity, we will consider the (asymmetric) 50-50 beam
splitters (see Fig.\ref{bs1}) which act on single photons in the
following way:
\begin{equation}\label{1ph}
|1\rangle \mapsto {{|3\rangle +|4\rangle}\over \sqrt 2} ,~~~~~~~~~~
 |2\rangle \mapsto {{|4\rangle -|3\rangle}\over \sqrt 2}.
\end{equation}
If the incident beams in ports 1 and 2 are described by the
quadrature-wave function  $\Psi(x_1,x_2)$, then the beamsplitter
described by (\ref{1ph}) leads to the transformation \cite{Ulf}:
\begin{equation}\label{BS}
   \Omega (x_1,x_2) \mapsto
\Omega \left(\frac{x_4+x_3}{\sqrt{2}},\frac{x_4-x_3}{\sqrt{2}}
\right ).
\end{equation}

We start with the initial state $\psi(x_1)$ for the quadrature-wave
function of the input beam, $\phi(x_2, x_5)$ for that of the
(approximate) EPR-pair (the quantum channel) and initial ``ready"
states of two measuring devices:
\begin{equation}\label{psiin}
   |\Psi\rangle = \int \psi \left( x_1 \right) |x_1\rangle dx_1 \int
   \phi \left( x_2 ,x_5 \right)  |x_2\rangle |x_5\rangle
 dx_2 dx_5 ~|ready\rangle_{_{M \hspace {-.05cm}D1}}
  |ready\rangle_{_{M \hspace{-.05cm} D2}}.
\end{equation}
Using Eq. (\ref{BS}), we have for the total state after the action
of the beam splitter:
\begin{equation}\label{psiin1}
 \int \psi\left(\frac{x_4+x_3}{\sqrt{2}}\right)
\phi \left(\frac{x_4-x_3}{\sqrt{2}},x_5 \right) |x_3\rangle
|x_4\rangle |x_5\rangle  dx_3 dx_4 dx_5 |ready\rangle_{_{M
\hspace{-.05cm} D1}} |ready\rangle_{_{M \hspace{-.05cm} D2}}.
\end{equation}
At this stage, $x_3$ and $p_4$ are measured and the appropriate
correction is applied to the state of the variable $x_5$. The wave
function in the $x$ representation is shifted by   ${\sqrt 2}x_3$,
and the wave function in the $p$ representation is shifted by
${\sqrt 2} p_4$. At this stage of our analysis we will not introduce
the ``collapse" of the quantum measurement, but continue to include
the measuring device in the description of the total state. The
shift in $x$ leads to the following transformation:
\begin{equation}\label{psiin2}
\int \psi\left(\frac{x_4+x_3}{\sqrt{2}}\right)
\phi\left(\frac{x_4-x_3}{\sqrt{2}},x_5-  {{2x_3}\over
{\sqrt{2}}}\right) |x_3\rangle |x_3\rangle_{_{M \hspace{-.05cm} D1}}
|x_4\rangle |x_5\rangle dx_3 dx_4 dx_5 |ready\rangle_{_{M
\hspace{-.05cm} D2}}.
\end{equation}
To see the effect of the shift in $p$, we apply a Fourier transform
in $x_4$ and then multiply the function of $x_5$ by $e^{i{\sqrt
2}x_5 p_4}$:
\begin{equation}\label{psiin3}
 \int {{e^{-i (x_4 -{\sqrt 2}x_5)p_4}}\over{\sqrt{2\pi}}}
 \psi\left(\frac{x_4+x_3}{\sqrt{2}}\right)
\phi\left(\frac{x_4-x_3}{\sqrt{2}},x_5- {{2x_3}\over
{\sqrt{2}}}\right) |x_3\rangle |x_3\rangle_{_{M \hspace{-.05cm} D1}}
|p_4\rangle |p_4\rangle_{_{M \hspace{-.05cm} D2}} |x_5\rangle dx_3
dx_4 dx_5 dp_4
\end{equation}
In  the limiting case the an ideal EPR pair,
\begin{equation}\label{delta
}
   \phi(x_2,x_5)=\delta(x_2-x_5),
\end{equation}
and the state (\ref{psiin3}), after the integration on $x_4$, has
the form:
\begin{equation}\label{psiin4}
 \int {{e^{i x_3 p_4}}\over{\sqrt{2\pi}}} |x_3\rangle  |p_4\rangle
|x_3\rangle_{_{M \hspace{-.05cm} D1}} |p_4\rangle_{_{M
\hspace{-.05cm} D2}} dx_3 dp_4\int \psi(x_5) |x_5\rangle dx_5.
\end{equation}
Thus, we have the desired teleportation of the wave function from
mode 1 to mode 5. Of course, no information about the teleported
state remains in the measuring devices.

In the process, we assumed ideal homodyne detectors and an ideal EPR
source. The major difficulty is the creation of the EPR source. An
approximate EPR state is obtained by shining beams of squeezed light
on a beamsplitter (Fig.\ref{telepo}). The light in input mode $a$
should be highly squeezed in the $x$ quadrature and the light on the
input mode $b$ should be highly squeezed in $p$. The input beams are
well approximated by Gaussians:
\begin{equation} \label{Gauss}
     {1\over {\pi^{1\over 4}
     \sqrt{\sigma_a}}}{e^{-{{x^2_a}\over{2\sigma^2_a}}}}
     ,~~~~~~~~
     {1\over {\pi^{1\over 4}
     \sqrt{\sigma_b}}}{e^{-{{x^2_b}\over{2\sigma^2_b}}}}
     ,
\end{equation}
where $\sigma_a$ is very small and $\sigma_b$ is very large. We will
require:
\begin{equation}\label{oomag}
    \sigma_a \ll 1/|p_1|,~~~ \sigma_b \gg |x_1| .
\end{equation}for all probable values of $x_1,~~p_1$.
    For input beams (\ref{Gauss}), instead of the ideal EPR state we will get
\begin{equation}\label{p}
  \phi(x_2,x_5)=
  \frac{1}{\sqrt{\pi \sigma_a \sigma_b}}e^{-(\frac{x_2-x_5}{2\sigma_a})^2}
  e^{-(\frac{x_2+x_5}{2\sigma_b})^2} .
\end{equation}

 In order to get a feel for the distortion during the teleportation, we
will consider two separate cases: one in which only the squeezed
light in port $a$ is not ideal and one in which only the squeezed
light in $b$ is not ideal. In the first case
\begin{equation}\label{p01}
\phi(x_2,x_5)=
  \frac{1}{\pi^{1/4}\sqrt {\sigma_a}}e^{-(\frac{x_2-x_5}{2\sigma_a})^2}.
\end{equation}
Then, the
  final state of the teleportation procedure (up to normalization)
  obtains the form:
  \begin{equation}\label{tel10}
    \int e^{ix_3 p_4}
     e^{-\left(\frac{x_5-v}{2\sigma_a}\right)^2}
       e^{-i\sqrt{2}p_4(v-x_5)}\psi(v)
       |x_3\rangle |p_4\rangle|x_5\rangle
    |x_3\rangle_{_{M \hspace{-.05cm} D1}}|p_4\rangle_{_{M \hspace{-.05cm} D2}}
     dx_3 dp_4 dx_5 dv.
  \end{equation}
Now, there is a partial entanglement between the system with the
teleported state and the measuring devices. Let us look at a
particular outcome of the measurement in mode 4, $p_4$. This
eliminates the entanglement. Then, the final teleported state  (up
to normalization) is:
\begin{equation}\label{conv}
\psi_{tel}(x_5)=\int e^{-i\sqrt{2}p_4(v-x_5)}
e^{-\left(\frac{x_5-v}{2\sigma_a}\right)^2}\psi(v)dv .
\end{equation}
This is just a convolution of the input function with a (real)
Gaussian multiplied by an (imaginary) exponent. If the Gaussian is
narrow and we can neglect the distortion due to the exponent, the
convolution yields approximately the input wave function, i.e., we
obtain teleportation with good fidelity.

The distortion due to the exponent depends on the value of $p_4$. In
order to estimate it, we can write an operator equation similar to
those that appear in the original continuous teleportation paper
\cite{mytele}:
\begin{equation}\label{eql}
    p_4={{p_1}\over\sqrt2}+{{p_a +p_b}\over 2},
\end{equation}
where $p_1$ is the `momentum' of the input mode and $p_a$ and $p_b$
are  the `momenta' of the input modes of the EPR source. For our
input state we have:
\begin{equation}\label{pa}
\langle p_a \rangle = 0,~~~ \Delta p_a = 1/\sigma_a, ~~~p_b=0.
\end{equation}
Taking into account the first of the requirements (\ref{oomag}), we
see that for probable outcomes of the measurement of $p_4$, $|p_4|
\lesssim \frac{1}{\sigma_a}$. Since we consider a narrow Gaussian
such that $\psi(x)$ is nearly constant on an interval of length
$\sigma_a$, the exponent does not lead to a significant distortion
and we obtain:
\begin{equation}
\psi_{tel}(x_5)\simeq \psi(x_5).
\end{equation}

In the other case we had mentioned above, namely that only the
squeezing in mode $b$ is not ideal, the `EPR' state is given by:
\begin{equation}
\phi(x_2,x_5)=\delta(x_2-x_5)\frac{e^{-\left(\frac{x_2+x_5}{2\sigma_b}\right)^2}}
{\pi^{1/4}\sqrt{\sigma_b}}.
\end{equation}
The final state after the teleportation procedure is now (up to
normalization):
\begin{equation}\label{tel2}
    \int e^{ix_3 p_4}
     e^{-\left(\frac{x_5-\sqrt{2}x_3}{\sigma_b}\right)^2} \psi(x_5)
       |x_3\rangle |x_3\rangle_{_{M \hspace{-.05cm} D1}}
       |p_4\rangle |p_4\rangle_{_{M \hspace{-.05cm} D2}}|x_5\rangle
     dx_3 dp_4 dx_5.
  \end{equation}
Again, there is a partial entanglement between the system with the
teleported state and the measuring devices. Let us look at a
particular outcome of the measurement in mode 3, $x_3$. This
eliminates the entanglement. Then, the final teleported state  (up
to normalization) is:
\begin{equation}\label{mult}
\psi_{tel}(x_5)=\psi(x_5)e^{-\left(\frac{x_5-\sqrt{2}x_3}{\sigma_b}\right)^2}.
\end{equation}

For the distortion to be small, we need the Gaussian to be
approximately constant over the interval where $|\psi|$ is
significant. Let us denote the length of this interval by $l$ which,
according to our second choice in (\ref{oomag}), is much smaller
than $\sigma_b$. The condition is then:
\begin{equation} \label{condi}
x_3\ll \sigma_b^2/l.
\end{equation}
To estimate the range of the outcomes of measurement results of
$x_3$, we can write (compare with (\ref{eql})):
\begin{equation}
x_3=\frac{x_1}{\sqrt{2}}+\frac{x_a+x_b}{2}.
\end{equation}
In this case, the input state of the EPR source is characterized by
\begin{equation} \label{esti}
x_a=0,~~~\langle x_b \rangle =0,~~~\Delta x_b=\sigma_b,
\end{equation}
which together with (\ref{oomag}), ensures that condition for good
fidelity teleportation (\ref{condi}) is satisfied.

We have seen two cases where simple estimation of teleportation
distortion was possible: in one case the operation was essentially
convolution with a narrow Gaussian, and in the other case the
operation was essentially multiplication by a wide Gaussian. Note,
that we could consider the process of teleportation in the $p$
representation and then we would have the same explanations, but
multiplication in the first case  and convolution in the second.

 We
have shown that either of the choices
\begin{equation}\label{choi}
\sigma_a \ll 1/|p_1|, \sigma_b=\infty ~~~ {\rm or}~~~ \sigma_a=0,
\sigma_b \gg |x_1|
\end{equation}
ensures that the distortion shall be small. In a real experiment,
both input beams for the EPR source exhibit, of course, finite
squeezing (\ref{Gauss}). Then, we can see from (\ref{psiin3}) that
the
  final state of the teleportation procedure (up to normalization)
  is:
  \begin{equation}\label{tel1}
    \int e^{ix_3 p_4}
     e^{-\left(\frac{x_5-v}{2\sigma_a}\right)^2}
     e^{-\left(\frac{x_5+v-\sqrt{2}x_3}{2\sigma_b}\right)^2}
       e^{-i\sqrt{2}(v-x_5)p_4}\psi(v)
       |x_3\rangle |x_3\rangle_{_{M \hspace{-.05cm} D1}}|p_4\rangle
    |p_4\rangle_{_{M \hspace{-.05cm} D2}}|x_5\rangle dx_3 dp_4 dx_5 dv
  \end{equation}
Given particular outcomes of the measurements of $p_4$ and $x_3$,
the final teleported state (\ref{tel1}) (up to normalization) is:
\begin{equation}
    \psi_{tel}(x_5)=\int
    e^{-\left(\frac{x_5-v}{2\sigma_a}\right)^2}
     e^{-\left(\frac{x_5+v-\sqrt{2}x_3}{2\sigma_b}\right)^2}
       e^{-i\sqrt{2}(v-x_5)p_4}\psi(v)dv \label{limit1}.
\end{equation}
 The analysis of this expression is more complicated, but condition
(\ref{oomag}) is sufficient to ensure that the function will be
teleported with little distortion, as will be seen in the computer
simulation presented in the following section.

\section{A Numerical Simulation}

In this section we show numerically the impact of each parameter in
the teleportation. We use as our input wave function the silhouette
of a woman Fig.(\ref{original}). We would like to stress that the
choice of a human figure is intended to aid in the visual assessment
of the distortion and is not, of course, supposed to suggest that
teleportation of humans is feasible.

The characteristics of the spatial wave function are:
\begin{equation}\label{charac}
\langle x_1 \rangle=50, ~~~\Delta x_1 =28,~~~~ l =100,
\end{equation}
where we have chosen units such that $\hbar =1$. Then, the computer
calculations show that the characteristics of the wave function in
momentum space are:

\begin{figure}[t]
\begin{center}
\includegraphics[width=0.7\textwidth]{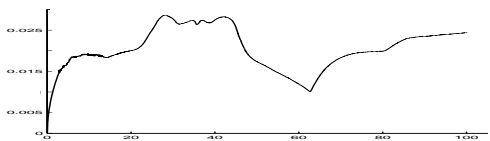}
\end{center}
\caption{The original wave function $\psi (x_1)$.} \label{original}
\end{figure}

\begin{figure}[b]
\begin{center}
\includegraphics[width=0.8\textwidth]{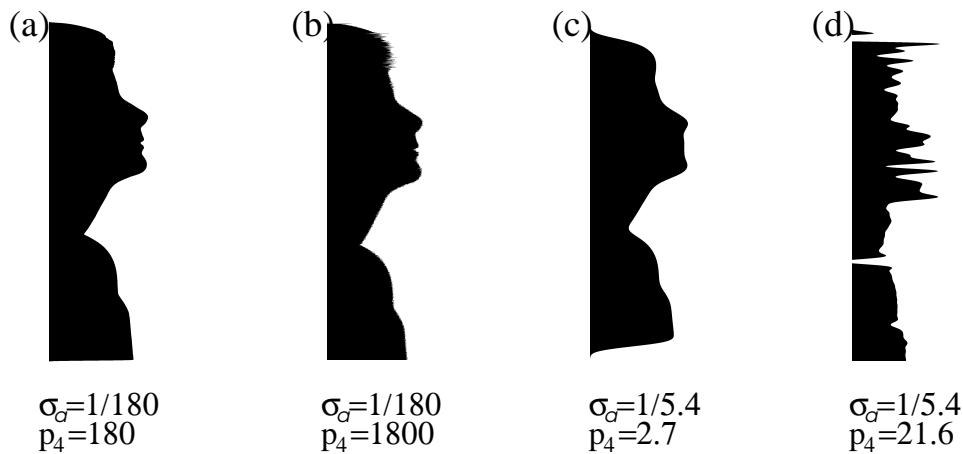}
\end{center}
\caption{Teleported silhouette with finite squeezing in $a$ and
ideal squeezing in $b$} \label{table2}
\end{figure}

\begin{equation}\label{characp}
\langle p_1 \rangle=0, ~~~\Delta p_1 = 18.
\end{equation}

First, we consider the case of infinite squeezing in port $b$ and
finite squeezing in port $a$. In Fig. \ref{table2}a we show the
results of the simulation for strong squeezing,
$\sigma_a=1/180(=1/10\Delta p_1$) and a probable outcome of the
measurement of the momentum in mode 4, $p_4=180(=1/\sigma_a$). Fig.
\ref{table2}b shows the results for the same strong squeezing, but a
rare large outcome of the measurement of the momentum, $p_4=1800 (=
10/\sigma_a)$. Fig. \ref{table2}c shows the results for weak
squeezing, $\sigma_a=1/5.4(=1/0.3\Delta p_1)$ and probable outcome of the
measurement of the momentum $p_4=2.7(=0.5/\sigma_a)$. Finally, Fig.
\ref{table2}d shows the results for weak squeezing,
$\sigma_a=1/5.4(=1/0.3\Delta p_1)$ and rare large outcome of the
measurement of the momentum, $p_4=10.8(=2/\sigma_a)$.

\begin{figure}
\begin{center}
\includegraphics[width=1\textwidth]{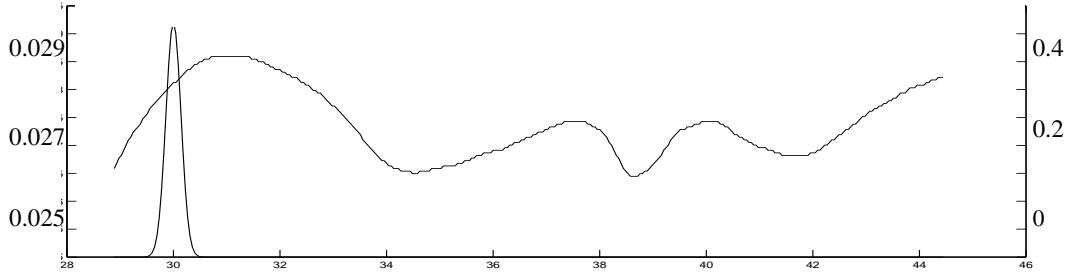}
\end{center}
\caption{The function $\psi(v)$ and (the real part of the)
convolution kernel of Eq. (\ref{conv2})($x_5=30$) corresponding to Fig.
\ref{table2}c. On the left is the scale of the wave function and on
the right, that of the kernel.} \label{char2}
\end{figure}

We see in Fig. \ref{table2}a that with strong squeezing and not too
large a value of $p_4$, the wave function is teleported without
significant distortion.
Fig. \ref{table2}b shows that for strong squeezing, but an
improbable large value of $p_4$ leads to distortion of the regions
which require large momenta. Fig. \ref{table2}c shows that weak
squeezing causes smoothing out of small details of the wave function
and large $p_4$ and weak squeezing lead to complete distortion of
the teleported wave function.

Let us analyze, for example, smoothing out of small details of the
wave function which we have seen in Fig. \ref{table2}c. The
teleported wave function is given by the convolution ({\ref{conv})
which, with the parameters $1/\sigma_a=5.4$ $p_4=2.7$ , reads:
\begin{equation}\label{conv2}
\psi_{tel}(x_5)=\int e^{-i3.8(v-x_5)}
e^{-\left(\frac{x_5-v}{0.37}\right)^2}\psi(v)dv .
\end{equation}
In Fig. \ref{char2} we draw the terms of the convolution
(\ref{conv2}). The graph shows only the real part. The imaginary
part is an order of magnitude smaller then the real part. The
multiplication of the Gaussian by the exponent has small effect,
since the width of the Gaussian is smaller than most details of the
wave function $\psi(v)$. Thus, the result of the convolution is
close to the original wave function, yet some details do disappear,
like the mouth.

\begin{figure}[b]
\begin{center}
\includegraphics[width=0.8\textwidth]{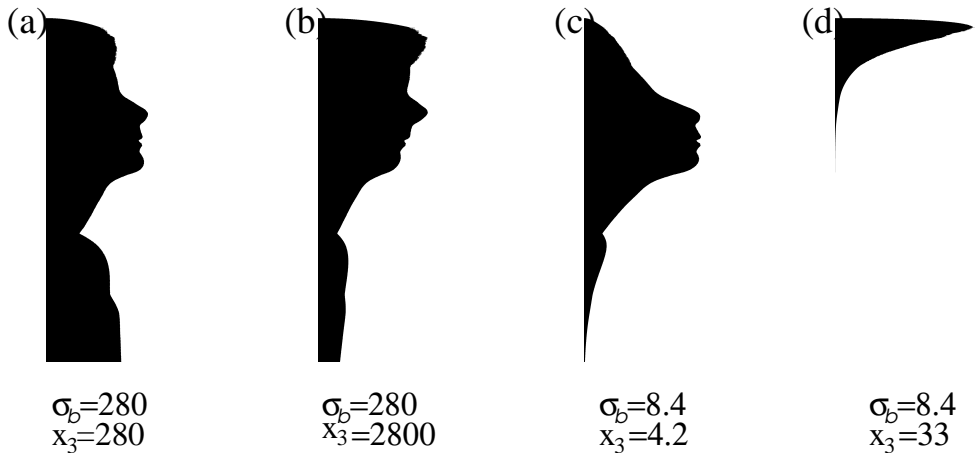}
\end{center}
\caption{Teleported silhouette with finite squeezing in $b$ and
ideal squeezing in $a$} \label{table1}
\end{figure}

\begin{figure}
\begin{center}
\includegraphics[width=1\textwidth]{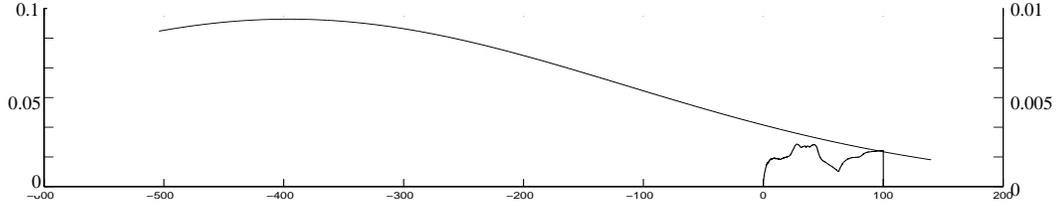}
\end{center}
\caption{The function $\psi(v)$ and the Gaussian which are the terms
of the product  (\ref{mult2}) corresponding to Fig. \ref{table1}a.
On the left is the scale of the wave function and on the right, that
of the Gaussian.} \label{char1}
\end{figure}

The second case is infinite squeezing in port $a$ and finite
squeezing in port $b$. In Fig. \ref{table1}a we show the results of
the simulation for strong squeezing, $\sigma_b=280(=10\Delta x)$ and
a probable outcome of the measurement of position,
$x_3=280(=\sigma_b)$. Fig \ref{table1}b shows the results for the
same  good squeezing, but rare large outcome of the measurement of
position,  $x_3=2800(= 10\sigma_b)$. Fig. \ref{table1}c shows the
results for weak squeezing, $\sigma_b=8.4(=0.3\Delta x)$ and
probable outcome of the measurement of position $x_3=4.2(=
0.5\sigma_b)$. Fig. \ref{table1}d shows the results for weak
squeezing, $\sigma_b=8.4$ and rare large outcome of the measurement
of the momentum, $x_3=33( =4\sigma_b)$.

\begin{figure}[b]
\begin{center}
\includegraphics[width=0.6 \textwidth]{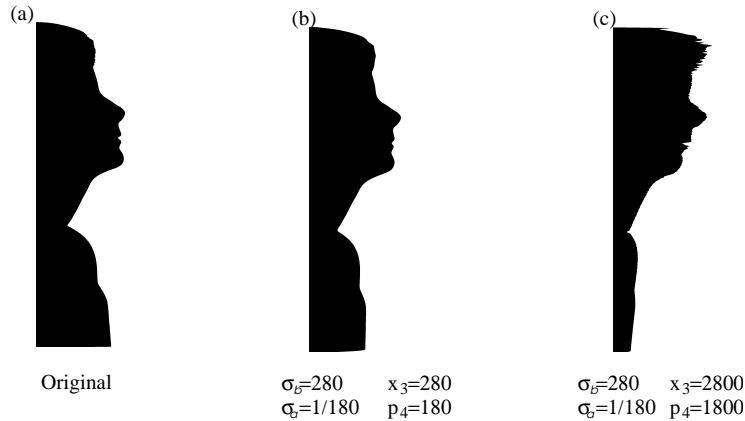}
\end{center}
\caption{Teleported silhouette with finite squeezing in $a$ and $b$}
\label{total}
\end{figure}

We see in Fig. \ref{table1}a that with strong squeezing and not too
large a value of $x_3$, the wave function is teleported without
significant distortion. It is the product (\ref{mult}) of the
original function with a wide Gaussian, which for the above
parameters reads:
\begin{equation}\label{mult2}
\psi_{tel}(x_5)=\psi(x_5)e^{-\left(\frac{x_5+400}{280}\right)^2}.
\end{equation}
Fig. \ref{char1} shows the wave function and the Gaussian of the
product (\ref{mult2}). We can see that the Gaussian is nearly
constant over the support of the wave function, thus causing little
distortion.

Fig. \ref{table1}b shows that strong squeezing,
but improbable large value of $x_3$ lead to distortion of the
relative amplitude. Fig. \ref{table1}c shows that weak squeezing and
small $x_3$ yield relatively faithful teleportation of a small part
of the wave function and distortion (elimination) of other parts.
Again, weak squeezing together with large $x_3$  lead to complete
distortion of the teleported wave function.

Finally, we compute the teleported wave function when both input
states are not ideal. Fig. \ref{total}b shows that strong squeezing
$\sigma_b=280(=10\Delta x),~\sigma_a=1/180(=1/10\Delta p)$ and probable
outcomes $p_4=180(=1/\sigma_a), x_3=280(=\sigma_b)$ together lead to a
very good fidelity, while large improbable outcomes
$p_4=1800(=10/\sigma_a), x_3=2800(=10\sigma_b)$ lead to significant
distortion, see Fig. \ref{total}c.

The discussion of the physical meaning of teleportation and the
analysis of continuous variables teleportation experiment which we
performed here provides an intuitive picture of the process and
helps to understand the effect of various parameters on the fidelity
of teleportation. We hope it will be useful for designing better
teleportation experiments.

It is a pleasure to thank Shai Machnes for helpful discussions. This
research was supported in part by grant 62/01 of the Israel Science
Foundation. N.E. gratefully acknowledges support from the Robert A.
Welch Foundation, grant no. A-1218.



\begin{thebibliography}{11}
\bibitem{Ppriv}
Asher told me that the name should be ``energy and momentum
exchange-free measurements".
\end{thebibliography}
\end{document}